# Unraveling the Evolution of Defectors in Online Business Games

Sanat Kumar Bista, Keshav P. Dahal, Peter I. Cowling and Bhadra Man Tuladhar

*Abstract*— Anonymous online business environments have a social dilemma situation in it. A dilemma on whether to cooperate or Defect. Defection by a buyer to seller and/or seller to buyer might give each a better profit at the cost of the loss of other. However, if these parties were to interact in future too, a bad past reference might prevent cooperative actions, thus depriving each other from a better gain. The anonymity of the players and an absence of central governing body still make this environment tempting for the defectors. What might be the evolutionary behavior of defectors in such environment? How could their increasing population be controlled? It is these two questions basically that we attempt to address in this research work. A genetic algorithm based spatial iterated prisoner's dilemma (SIPD) environment has been used to simulate the experiments. A case where compensation for the *looser* is provided by the system is modeled and analyzed through experiments. Our results show that compensation can be useful in decreasing defective population in the society, however, this might not be enough for the evolution of a cooperative and reliable society of trustworthy players.

## I. INTRODUCTION

Businesses are typically characterized by a motive to maximize profit. To a seller, maximization of profit allows greater benefit and also serves its sustainable growth in a competitive market environment. Similarly, a buyer would also look for a suitable bargain in the cost price to maximize its gain. When a cooperative seller interacts with a cooperative buyer, both of the *players* (we would use this term *player* to denote a buyer or seller in general) would end up in a happy transaction.

The seller makes its profit while the buyer gets the goods in a better price, thus making the transaction beneficial to both of them. In a setting where they are dealing in a credit transaction, if one of them turns out to be selfish, the cooperator is badly defected, while the selfish one benefits the most. In such an environment, any rational player finds incentives for its selfishness and also enjoys a lesser risk over cooperation which might lead to its defeat in case the other player defects. Thankfully, in the real world, this situation is eased by the presence of governing bodies and the possibility of physical identification of players. Further, selfish behavior of a player can be at the cost of its credibility which is an important thing for a player to maintain in order to ensure a renewed interest of the other party in future transactions. Cooperation however might not be an easy thing to achieve in an online business environment (generalized term for electronic markets) where anonymity of a player might aid in making the situation manipulable by the strategic ones. The types of environments that we consider here are typically characterized by the lack of a central governing body which would verify the identity of the players and maintain 'a law and order situation' in it. Today's popular e-Business sites like eBay and the Peer to Peer networks in their extreme form can be an example of this type of environment. EBay maintains centrally the data of its members and lists their credibility by providing publicly their reputation information [1]. These information to certain extent, help the player in perceiving security in transaction, but still in absence of a physical security, strategic players can earn higher payoff for their selfish and deceptive act. How then would cooperation evolve in online business scenario? It is this interesting aspect of cooperation evolution that we are investigating through our experiments. In this paper, we report the outcomes of our investigation into how and under what conditions cooperation might evolve in an online business environment which lacks a central governing body.

The remaining part of this paper is structured as follows: Section II gives background information to IPD and On-line business game and also reviews some of the related works in this area. The details of the system are presented in Section III. Section IV lists some experimental results and analysis. Conclusion and future works in this area is highlighted in Section V.

Sanat Kumar Bista is with the MOSAIC Research center at the University of Bradford, BD7 1DP, UK as a PhD Research Student. (Corresponding author: phone: +44(0)1274233948, fax:+44 (0) 1274 233920; e-mail: s.k.bista@ bradford.ac.uk).
Keshav P. Dahal is with the MOSAIC Research Center at the University of Bradford, BD7 1DP, UK as a Senior Lecturer. (e-mail: k.p.dahal@bradford.ac.uk).
Peter I Cowling is with the MOSAIC Research Center at the University of Bradford, BD7 1DP, UK as a Professor. (e-mail: p.i.cowling@bradford.ac.uk).
Bhadra Man Tuladhar is with the Department of Natural Sciences (Mathematics) at Kathmandu University, Nepal, as a Professor. dean_sci@ku.edu.np).

## II. IPD AND ON-LINE BUSINESS GAMES

We relate the situation in our online business environment to that in a Prisoner's Dilemma Game. Prisoner's dilemma like puzzle structures were first devised by Merrill Flood and Melvin Dresher in 1950, while the term "Prisoner's Dilemma" and the version with prison *sentences* as *payoffs* is credited to Albert Tucker [2]. Further description with interesting cases from society was given by Axelrod in [3] and the evolution of strategies in an Iterated Prisoners Dilemma (IPD) environment described by the same author in [4]. In a prisoner's dilemma game setting, there are two prisoners held for some crime and being interrogated separately. They share a dilemma as they don't get to see each other and one does not know what the other is going to say about the crime. Each prisoner has an option to give evidence against its partner by defecting him, or to cooperate by holding the evidence. Three particular situations can arise in this case:

i. Both of them can cooperate by holding evidence, thus the judge can have a less doubt over their guilt and decide for a relatively shorter imprisonment (Say 1 year)
ii. Both of them can defect by providing evidence against each other, thus making it easy to reveal their involvement. In this case the judge might decide for a longer imprisonment for both of them (say 3 years).
iii. One of them can defect while the other cooperates. In this case the defector might be set free while the cooperator being proven guilty could be given an even longer sentence (say 5 years)

This situation can also be expressed in a pay off matrix as in Table I below:

TABLE I
PAYOFF MATRIX FOR A TYPICAL PD GAME

|  |  | Prisoner B | |
|---|---|---|---|
|  |  | Cooperate (C) | Defect (D) |
| Prisoner A | C | Punish A=1 yr  Punish B=1 yr | Punish A=5 yrs  Punish B=0 yrs |
|  | D | Punish A=0 yrs  Punish B=5 yrs | Punish A=3 yrs  Punish B=3 yrs |

The values for the pay off are typically called Reward for cooperation (R), Temptation to defect (T), Sucker's Payoff (S) for cooperating a defective player, and Punishment for mutual defection (P). For the dilemma to hold, the following condition must hold true [3]:

$$T > R > P > S \quad (1)$$

In the prisoner's example above these values correspond to the benefit that each would derive from the respective actions. There is further an interesting consequence if the game is being played for more than one round. Typically known as Iterated Prisoner's Dilemma (IPD), in such game the pay-off for two times reward is higher than the summed payoff for Temptation and sucker payoff. Thus an additional condition given below must also hold true in this case [3] [2].

$$2R > T + S \quad (2)$$

The situation in the Iterated Prisoner's Dilemma is also a representative to online business environment[5] [6]. If we consider an online trading environment where two completely anonymous buyer and seller are interacting, they share a dilemma on whether or not the other party would reciprocate its cooperation. If the seller sends the good and the buyer sends money worth the goods, both of them are rewarded and the transaction meets a happy ending. If either of them defects while the other cooperates, then the cooperator is badly *hurt* in the transaction and the defector scores highest. If both defect each other then each scores a payoff better than *sucker's* score (and interestingly equal to reward in value as both posses the money and goods with them), but is nevertheless going to be a bad reference for future transaction. Further, a payoff for a both defect scenario cannot be considered equal to Reward, as there has been no transaction at all. A *zero* return for no transaction is what the payoff in this case would be.

A major difference this situation has with the traditional IPD tournaments is that, here it is not necessary for each player to play with every player in the society. A variation of IPD called Spatial IPD represents this scenario better. In spatial IPD the players are arranged in some "geographical" positions and most of the interactions take place between the neighboring players[2]. Such arrangements exhibit games in clusters thus providing natural representation of our problem [7]. As an example, in eBay like setting, sales items are put under several categories and each category has some defined sets of sellers who interact with buyers sharing similar interest. If we were to measure an overall evolution of cooperation, these individual categories of players are the source for it. It is in this environment that our research focuses on investigating the evolution of cooperation. The detail of the model is described in Section III.

In one of the most remarkable works on the evolution of cooperation Axelrod in [3] presents an optimism to cooperation by addressing fundamental realities of human nature. Starting from basic questions like when should a person be cooperative, and when selfish, his work tries to demystify the complexity of cooperation in humans by relating it to a Prisoner's dilemma situation. Based on Prisoner's dilemma again, Axelrod in [4] conducted a tournament of strategies in which he used Genetic Algorithms to identify a fittest and an evolutionarily stable strategy. Issues relating to the evolution of cooperation, Trust and Reputation have been addressed by various authors in [5], [6] [8], [9].

Aberer in [5] outlines the complexity of Trust and Reputation and discusses different approaches to computing trust and reputation. The authors have considered evolutionary approach as one of the many popular approaches that game theorists have been using. In [6] the authors have presented a social mechanism of reputation management in electronic communities. In their discussion around electronic communities, the authors have described the Prisoner's dilemma situation in it. In a related work Janssen in [8] has studied the role of reputation scores in the evolution of cooperation in online e-commerce sites. The author discusses whether or not reputation alone can be meaningful in evolving a cooperative society. The paper concludes that high level cooperation is not only possible with reputation scores. The author investigates the work in a one-shot prisoner's dilemma like environment. In [9], the author has studied the impact of providing incentive to cooperation as a side payment scheme to make it rational for the agents to be cooperative. The proposed model is described using an Iterated Prisoners' Dilemma scenario.

### III. THE SYSTEM

*A. Business Game Models*

The anonymous business game between two players, a buyer and a seller described in the introduction section can also be expressed in terms of a payoff matrix as in Table I. Let $\gamma$ be the price of goods that is being purchased by a buyer from some seller. The amount of money the buyer has to pay the seller is the price of the goods, thus in this case this also equals $\gamma$ in value. The gains of these two players in four different possible action sequences are summarized by the matrix in Table II below:

TABLE II
PAYOFF MATRIX FOR A TYPICAL BUSINESS GAME

|  |  | Buyer | |
|---|---|---|---|
|  |  | **Cooperate (C)** | **Defect (D)** |
| Seller | C | $R_{seller} = \gamma$ <br> $R_{buyer} = \gamma$ | $S_{seller} = -\gamma$ <br> $T_{buyer} = 2\gamma$ |
| | D | $T_{seller} = 2\gamma$ <br> $S_{buyer} = -\gamma$ | $P_{seller} = 0$ <br> $P_{buyer} = 0$ |

The notations T, R, P and S have the same meaning as in (1). The relationships between the values here are still conformant to the inequalities in (1) and (2). Thus this situation imposes a *Strong Dilemma* in the players.

Another typical business game is an eBay like scenario. We take eBay as a representative of online auction environments and there might be others in its class offering similar services. This setting is different from the one described in Table I, in a way that eBay has buyer and seller *protection programs* which compensates to some extent the loss incurred, if any [10]. Let $\delta$ be the factor with which losses are compensated. Including this compensation scheme the new payoffs for the game are given in Table III below:

TABLE III
PAYOFF MATRIX FOR A *COMPENSATED BUSINESS* GAME

|  |  | Buyer | |
|---|---|---|---|
|  |  | **Cooperate (C)** | **Defect (D)** |
| Seller | C | $R_{seller} = \gamma$ <br> $R_{buyer} = \gamma$ | $S_{seller} = -\gamma + \delta$ <br> $T_{buyer} = 2\gamma$ |
| | D | $T_{seller} = 2\gamma$ <br> $S_{buyer} = -\gamma + \delta$ | $P_{seller} = 0$ <br> $P_{buyer} = 0$ |

The compensation factor introduces a new relationship in the payoff inequalities described in (1) and (2) above. While the exact new condition is a factor of $\delta$ in reality, this addition of $\delta$ nevertheless changes the *Strong Dilemma* into a *Weak Dilemma* situation. The particular change is given in the equation below:

$P \geq S$ (3)

This small change introduces interesting consequence, as it increase the confidence of the player in acting cooperatively. In a real world scenario, if there is a third party protection for a transaction, the confidence of a buyer or seller increases, and this increase is proportional to the fraction of compensation provided.

We simulate the evolution of cooperation in each of the Cases (*Strong Dilemma,* and *weak dilemma*) described above and present the results in next section.

*B. Simulation Model*

At the highest level of abstraction, our simulation model consists of a two dimensional spatial grid of players on a square surface with overlapping edges. Each player has eight neighbors with whom it plays the cooperation defection game. Players are characterized by strategies for a 3-memory game. This means that each player while interacting with its neighbor will keep track of past three games. An action on what to do next will be guided by the particular strategy that matches the past three histories of interactions that each player maintains. Corresponding moves for initial interactions until the three games are played are also encoded in player strategies. Figure 1 below shows a sample 5x5 square grid with 25 players in it. The highlighted player P7 plays games for a specified number of times with its eight neighbors who are P1, P2, P3, P8, P6, P13, P12, and P11. The games mentioned here represent the business interactions described in the Introduction section and section III A above.

| P0 | **P1** | **P2** | **P3** | P4 |
|---|---|---|---|---|
| P5 | **P6** | P7 | **P8** | P9 |
| P10 | **P11** | **P12** | **P13** | P14 |
| P15 | P16 | P17 | P18 | P19 |
| P20 | P21 | P22 | P23 | P24 |

Fig. 1. A 5X5 spatial grid with a player P7 highlighting its 8 neighbors

The actual evolution in the game is based on Genetic Algorithm. Genetic Algorithm (GA) as an intelligent search technique was introduced by computer scientist John Holland [11] [12] [4]. Our use of GA based evolution is based on, and inspired by, the original work of Axelrod in [4]. In the GA environment, a player strategy is represented by a chromosome as a fixed length representation of the possible actions for a three memory game. With two possible actions (Cooperate or Defect) that can be played by each of the two interacting players, the all possible moves are CC, CD, DC, and DD. A three memory strategy for these four possible moves needed 4x4x4=64 bit chromosome length. Axelrod in [4] used additional 6 bits to determine the first three moves. A variation of this was used by Errity in [13] and we are following the same scheme of additional bit encoding, in which 7 extra bits are used for encoding actions for the first three relative moves (relative to opponent moves). In this approach it is not required to encode an assumption of the pre-game history [13]. This makes the total chromosome length to be 71 bits with each locus outlining an action C or D to perform. The model when initialized consists of a specified number of players with random strategies. In each generation of evolution a pre-specified number of games are played between two players, at the end of which, the fitness of the chromosome of each player is evaluated by referring to the payoff values for the game. From player A's perspective, a CC action represents Reward Payoff, CD a Sucker's Payoff, DC a Temptation Payoff and DD a Punishment Payoff. Fittest strategies are selected for reproduction through crossover and mutation. During crossover, both the parent chromosomes are broken in at the same random point. Linear scaling as described in [12] of the fitness has been used to prevent premature convergence of the evolution. The results obtained for the cases in section III A by using this model is explained in the next section.

## IV. EXPERIMENTAL RESULTS AND ANALYSIS

### A. The Setup

The experiments were carried out for three different cases. In the first case (CASE I), the *strong dilemma situation* outlined in Table II was considered and next we focused on the *weak dilemma* situation outlined in Table III. In the later case we performed the experiments for two sub cases. The first (CASE IIA) being a situation were the loss compensation percentage was 50, and in the second (CASE IIB) this percentage was set to 100. Thus, for the first case $\delta = R/2$ and in second case $\delta = R$. The following parameters in table IV were used for all three experiments:

TABLE IV
SIMULATION PARAMETER

| Parameter | Value |
|---|---|
| Total number of Players | 2500 |
| Generations of Evolution | 1000 |
| Transactions per Generation | 200 |
| Crossover Probability | 0.98 |
| Mutation Probability | 0.01 |

The player population of 2500 was chosen to represent a good size of strategies. Simulations were limited to 1000 generations as at this point, the evolution was relatively stable. We had to be careful with the number of transactions as this number was proportional to the time taken by the simulation. However, a small value for transaction frequency might prevent the players from exploiting most of their *memory three* strategies. Axelrod in [4] had used an average run length of 151 for his strategy tournament. We took an intuitive value of 200 as a tradeoff between performance and fitness. The probabilities for crossover and mutation were kept fixed for all the transaction. Though a variance in these values might give different patterns of evolution, this analysis is currently is not included in our experiments.

Next set of parameters of the experiments were related to the payoff values which would be different for each of the cases outlined above. For this experiment we have assumed that all the goods being sold or purchased has the same value of £ 10, thus making reward for cooperation R=10. Based on this assumption, the other payoff values used for the three different cases are listed in the table V.

TABLE V
PAYOFF VALUES

| Payoff Category | CASE I | CASE IIA | CASE IIB |
|---|---|---|---|
| Reward (R) | 10 | 10 | 10 |
| Temptation (T) | 2R | 2R | 2R |
| Sucker's (S) | -R | -R+ δ | -R+ δ |
| Punishment (P) | 0 | 0 | 0 |
| Compensation (δ) | N/A | R/2 | R |

The initial population roughly consisted of around 80% of cooperative players and 20% of deceptive ones. As we have been trying to observe the trend of the evolution of defectors, their size was intentionally kept small initially so that a full span of their evolution could be visible in the results. In the definition, we consider a player to be cooperative, if more than 60% of its strategy guides cooperation to its action. Thus a defector has essentially less than 40% of cooperative actions in its strategy. We also consider in our experiments, a particular species of defector called *Top Defector*. This player has the most vindictive nature with more than 75% of its strategy guiding defective action to its opponent.

### B. Results and Analysis

From the cooperative evolution point of view, all the experiments gave depressive results. This can be easily derived from the graphs in Figure 2, as a graph for cooperative evolution would be complementary to these. To reach these results, the simulations for each of the cases listed in Table V were carried out for five times and an average taken to minimize the noise in evolution statistics. Fig 2a. Shows the evolution of defectors in a *strong dilemma situation*. This situation, as described earlier, is a representative of an anonymous online business environment. In a condition where there is no mechanism which can enforce cooperation, it is natural that defection prevails, as this would give success to the player. The population of defectors which started from a mere 17% sharply increased until it reached a maximum of 74% in the 724[th] generation, after which it maintained a population in between 60-70%. This trend reflects the high insecurity in transactions in these settings.

What might be the result if the ones who lost in transaction were compensated? Would it contribute in reducing the defectors in the society? The two graphs in Fig 2a and 2b have an answer to these questions. The results show that compensation certainly has negative impact on the growth of defectors and this impact is in some value proportional to the fraction of compensation. Fig 2b is a 50% compensation case. The result shows that the population of defectors in this setting has decreased as compared to the previous case, with a maximum population of only 55.16% in 275[th] generation. In case where a defected cooperator is returned full of the amount of goods or money, the cooperation evolution scenario was even better, with a record of only 45.70% of highest defectors in 233[rd] generation.

The three figures discussed above represented the overall population of defectors. The impact of compensation was sharply visible in a particular species of defector which we call here *Top Defector*. This species has a strategy which allows it to cooperate only 25% of time. The evolution of this species is considered by the graph in Fig 3. The population of this species was highest in the strong dilemma case and a lower and yet lower portion was seen in the other two cases. This suggests that the compensation scheme was successful in bringing down the population of this category of players. However, defective players are still prevalent in any of the cases examined above. The outcome of these experiments suggests that an even stronger model to weaken the dilemma is required in order to promote cooperative evolution in online business settings. Investigation of such techniques which can curb the defector population while increasing the size of cooperators is the future work of our research.

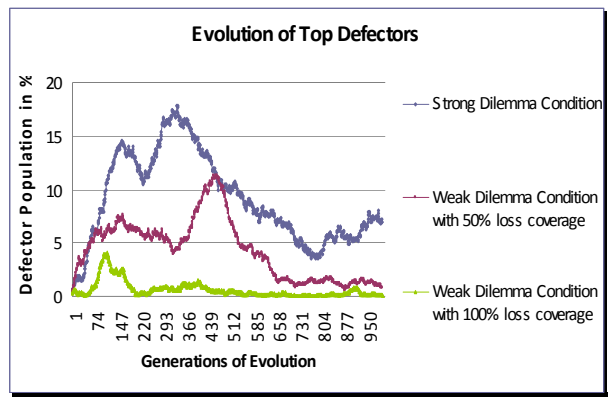

Fig 3. Evolution of Top Defectors in various conditions

## V. FUTURE WORK AND CONCLUSION

It was not originally our intention to cast the *"villains"* (defectors) in our work. We initially started with the trend in evolution of cooperators. Ironically, the model that we were using in our experiments (which in some way resembles the real world) gave depressive results for the evolution of cooperators. Insignificant impact of the model was seen in the evolution of cooperative players. This gave us a better insight into how *fragile* the situation was. We then turned our idea the other way round to see if these models had any impact on the defectors at all. The results showed that the models certainly had an impact on defectors and severely on the most selfish species which we named as *Top Defectors*. This is however not sufficiently reliable for online transactions; a conclusion that we have drawn from our experiments.

For the evolution of *safer societies* for transaction, we might need more powerful models which could attack on the *dilemma conditions* and make it rational for the players to cooperate. Compensation to the *looser* is one approach. An approach that could complement the need here might be to provide incentives to the cooperators. In real world online business scenarios however, it might not be rational to directly give a benefit to someone just because they acted nicely. Indirect reflection of incentives on their *reputation* might be a promising idea of rewarding that could lead to cooperative evolution. Our future work in this regard would be in identifying to what extent can reputation systems in real world contribute in the evolution of cooperation in society. We expect these results combined with the loss compensation scheme to give insightful feedback on the trustworthiness of players in online business environments.

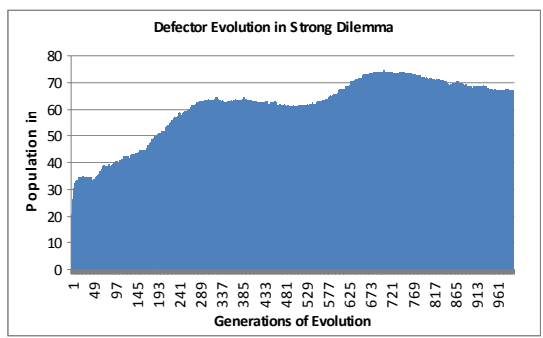

(a)

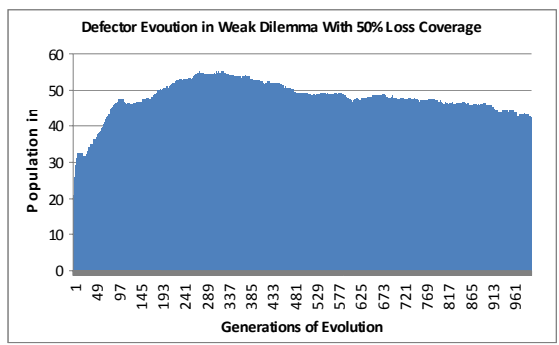

(b)

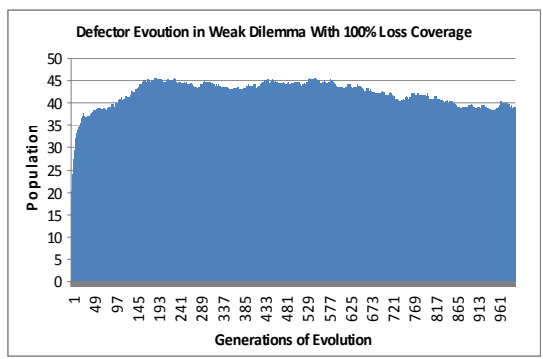

(c)

Fig 2. Evolution of Defective Players in (a) Strong Dilemma Conditions as in equation (1) and (2), (b) Weak Dilemma Condition with 50% Loss coverage, (c) Weak Dilemma Condition with 100% loss coverage